\documentclass[aps,pra,twocolumn,10pt]{revtex4-1}
\usepackage{graphicx,amsmath,amssymb}
\usepackage[utf8x]{inputenc}
\usepackage{url}
\usepackage{float}

\begin{document}

\title{Polarization and spatial coherence of electromagnetic waves in uncorrelated disordered media}

\author{Kevin Vynck}
\altaffiliation{Current address: Laboratoire Photonique, Num\'{e}rique et Nanosciences (LP2N), UMR 5298, CNRS - IOGS - Univ. Bordeaux, Institut d'Optique d'Aquitaine, 33400 Talence, France.}
\email{kevin.vynck@institutoptique.fr}
\author{Romain Pierrat}
\email{romain.pierrat@espci.fr}
\author{R\'{e}mi Carminati}
\email{remi.carminati@espci.fr}

\affiliation{Institut Langevin, ESPCI ParisTech, CNRS, 1 rue Jussieu, 75238 Paris Cedex 05, France.}

\date{\today}

\begin{abstract}
Spatial field correlation functions represent a key quantity for the description of mesoscopic phenomena in disordered media and the optical characterization of complex materials. Yet many aspects related to the vector nature of light waves have not been investigated so far. We study theoretically the polarization and coherence properties of electromagnetic waves produced by a dipole source in a three-dimensional uncorrelated disordered medium. The spatial field correlation matrix is calculated analytically using a multiple scattering theory for polarized light. This allows us to provide a formal description of the light depolarization process in terms of ``polarization eigenchannels'' and to derive analytical formulas for the spatial coherence of multiply-scattered light.
\end{abstract}

\maketitle

\section{Introduction}

Light transport in disordered media is characterized by a multiple scattering process which randomizes the direction, phase and polarization of the propagating waves~\cite{Akkermans2007, Sheng2010}. It is widely accepted that after many scattering events, polarization effects can be omitted in the description of the average light intensity. This justifies the predominant use of scalar theories for multiple light scattering in the literature. However, when the problem involves either short distances away from a polarized source or spatial correlation functions of the field on short scales, the vector nature of electromagnetic waves should, in principle, not be neglected. 

Most previous works on polarized light in disordered media have investigated the progressive transfer of energy from the co-polarized to the cross-polarized component of the scattered light with respect to the incident light. This light depolarization due to multiple scattering has been observed in coherent backscattering~\cite{vanAlbada1987, Rosenbluh1987} and time-resolved reflection experiments~\cite{Vreeker1989, MacKintosh1989a, Dogariu1997, RojasOchoa2004a}, where the contribution of short pathlength trajectories on the optical response of the medium is significant. Along similar lines, much attention has been given to magneto-optical effects (e.g. Faraday rotation) in disordered media \cite{Erbacher1993, Rikken1996}, where an external magnetic field affects reciprocity in light propagation and thus, interference phenomena. Different theoretical models for the multiple scattering of polarized light have been developed over the years, among which a perturbative approach for the field amplitude and intensity~\cite{Stephen1986, MacKintosh1988, Ozrin1992, vanTiggelen1996, Muller2002}, a transfer matrix approach for the light depolarization on scattering sequences~\cite{Akkermans1986, Akkermans1988, Martinez1994, Xu2005}, and a vector radiative transfer approach for Rayleigh scatterers~\cite{Amic1997} or under conditions of predominantly forward scattering (i.e. for systems containing large particles)~\cite{Gorodnichev2007}. All in all, the current knowledge about polarized light in disordered media can be summarized by the existence of a set of ``modes'' inherent to the vector nature of light and of a typical length scale governing the light depolarization process.

More recent years have witnessed a renewed interest in polarization-related phenomena in disordered media. For instance, it was suggested that important information about the specific morphology of disordered media could be retrieved by probing their optical response at mesoscopic scales~\cite{Haefner2010} or from the near-field speckle statistics measured at subwavelength distances from their surface~\cite{Apostol2003, Carminati2010}. Fluctuations of the local density of states (or similarly, the so-called $C_0$ correlation~\cite{Shapiro1999, vanTiggelen2006, Caze2010}) in disordered optical materials were shown to be driven to a large extent by near-field interactions between the source and the scatterers~\cite{Caze2010, Birowosuto2010, Sapienza2011}. Polarization effects are fundamental on such scales. Another example is the Anderson localization of light, which relies on wave interference between multiply-scattered waves~\cite{Lagendijk2009}. While theoretical models for multiple light scattering and localization generally deal with scalar waves~\cite{Kroha1993, Lagendijk1996, vanRossum1999, Akkermans2007, Sheng2010}, it was recently suggested that light localization in uncorrelated disordered media disappears when polarization effects are taken into account~\cite{Skipetrov2013}. These examples illustrate well the fact that the vector nature of light enters a wide range of modern problems in optics of disordered media and many aspects related to it remain to be explored.

A requisite quantity for the theoretical description of polarization-related phenomena in disordered media is the spatial field correlation matrix, or coherence matrix, defined as~\cite{Mandel1995, Brosseau1998}
\begin{equation}\label{eq:1}
W_{ij}(\mathbf{r},\mathbf{r}')=\langle E_i (\mathbf{r}) E_j^\star (\mathbf{r}') \rangle,
\end{equation}
where $i$ and $j$ are Cartesian coordinate indices, $\mathbf{r}$ and $\mathbf{r}'$ indicate two points in space, $^\star$ stands for complex conjugation, and $\langle .. \rangle$ denotes an ensemble average. The matrix $W_{ij}(\mathbf{r},\mathbf{r}')$, which has been at the core of research in beam optics and atmospheric turbulence for decades~\cite{Wang1979, Andrews2005}, provides statistical information on the field correlation \textit{after average over disorder realizations}~\footnote{By contrast, in a single disorder realization, the field at a point is always fully polarized (i.e., it has a well-defined polarization state) and the fields at two neighboring points are related in a way that is specific to the realization.}. The important notions of \textit{polarization} and \textit{spatial coherence} relate respectively to the correlation between orthogonal field components at one point ($i \neq j$ and $\mathbf{r}'=\mathbf{r}$) and between parallel field components at two different points ($i = j$ and $\mathbf{r} \neq \mathbf{r}'$). Although these two quantities are intimately linked, they constitute the most fundamental characteristics of electromagnetic fields~\cite{Dogariu2013b}. 

In this article, we investigate theoretically the polarization and spatial coherence of multiply-scattered light produced by a dipole source in a three-dimensional uncorrelated disordered medium. We derive analytical expressions for the spatial field correlation matrix using a multiple scattering theory for polarized waves. The theoretical framework is similar to that used in Refs.~\cite{Stephen1986, MacKintosh1988, Ozrin1992, Muller2002} but our work goes beyond this initial model, allowing us to derive new results. We calculate the eigenmodes that govern the diffusion of polarization as well as the characteristic length scales that describe energy propagation through these ``polarization eigenchannels''. This gives an important insight into the light depolarization process away from a dipole source. We also provide analytical formulas for the spatial light coherence, showing that the correlation function strongly depends on the orientation of the field components with respect to the two observation points ($\mathbf{r}$ and $\mathbf{r}'$). Our study therefore unveils fundamental properties of electromagnetic waves in disordered media, that are relevant to the understanding of polarization-related phenomena in complex systems and may have important outcomes for medical and material science applications, where optical imaging and spectroscopy techniques could benefit from polarization-resolved measurements~\cite{Tuchin2006, Emile1996, Jacques2000, Gorodnichev2012}.

The article is organized as follows. In Sec.~\ref{sec:2}, we present the theoretical model used to treat the multiple scattering of polarized light and describe the derivation of the spatial field correlation matrix. The derivation of the main results on the diffusion of polarization and on the spatial coherence of light in disordered media are presented in Sec.~\ref{sec:3}. We conclude in Sec.~\ref{sec:4} with some perspectives and ideas for future studies. The details of our calculations are given in the Appendices.

\section{Theoretical model}\label{sec:2}

\subsection{Multiple-scattering expansion}

We consider a monochromatic electromagnetic wave with free-space wavevector $k_0=\omega/c$, $\omega$ being the wave frequency and $c$ the speed of light, propagating in a three-dimensional disordered medium with dielectric function $\epsilon(\mathbf{r})$. The electric field $\mathbf{E}$ satisfies the vector propagation equation
\begin{equation}\label{eq:waveeq}
\nabla \times \nabla \times \mathbf{E}(\mathbf{r}) - k_0^2 \epsilon(\mathbf{r}) \mathbf{E}(\mathbf{r})  = i \mu_0 \omega \mathbf{j}(\mathbf{r}) ,
\end{equation}
where $\mathbf{j}(\mathbf{r})$ is a source distribution in the disordered medium. We assume that the disordered medium is non-absorbing and is described by a scalar dielectric function of the form $\epsilon (\mathbf{r})=1+\delta \epsilon (\mathbf{r})$, where $\delta \epsilon (\mathbf{r})$, the fluctuating part of the dielectric function, obeys white-noise Gaussian statistics~\footnote{The validity of this model is discussed in Appendix A2.1.4 (``The $\delta$ function potential'') of Ref.~\cite{Akkermans2007}.}
\begin{equation}\label{eq:disorder}
\langle \delta \epsilon (\mathbf{r}) \rangle = 0 \quad \text{and} \quad \langle \delta \epsilon (\mathbf{r}) \delta \epsilon (\mathbf{r}') \rangle = u \delta(\mathbf{r}-\mathbf{r}') ,
\end{equation}
with $u$ an amplitude that will be determined below.

The $i$-th component of the electric field in Eq.~(\ref{eq:waveeq}) is related to the dyadic Green function $G_{ik}$ as
\begin{equation}\label{eq:field-green}
E_i(\mathbf{r}) = i \mu_0 \omega \int G_{ik}(\mathbf{r},\mathbf{r}') j_k (\mathbf{r}') d\mathbf{r}',
\end{equation}
where implicit summation of repeated indices is assumed. For statistically isotropic, translational-invariant media and assuming that the scatterers are in the far-field of each other (i.e. near-field interactions are neglected), one shows (through Dyson's equation) that the average dyadic Green function in reciprocal space is given by~\cite{Tai1994}
\begin{equation}\label{eq:greentensor}
\langle G_{ik} (\mathbf{q}) \rangle = (\delta_{ik} - \hat{q}_i \hat{q}_k) \langle G (\mathbf{q}) \rangle ,
\end{equation}
with $\langle G (\mathbf{q}) \rangle = \left( k_0^2 - q^2 - \Sigma(\mathbf{q}) \right)^{-1}$, the scalar average Green function and $\Sigma$ the self-energy, which contains the sum of all multiply-scattered events. To order $(k_0 \ell)^{-1}$, where $\ell$ is the scattering mean free path, one finds that $\Sigma(\mathbf{q}) \simeq -i k_0/\ell$, thereby expressing the attenuation of the field due to scattering, and $u=6\pi/k_0^4 \ell$~\cite{Akkermans2007,Sheng2010}.

The spatial field correlation matrix $W_{ij}(\mathbf{r},\mathbf{r}') = \langle E_i(\mathbf{r}) E_j^\star (\mathbf{r}') \rangle$ can be derived from standard diagrammatic calculations~\cite{Akkermans2007, Sheng2010}, and takes the form of a Bethe-Salpeter equation
\begin{eqnarray}\label{eqn:bethesalpeter-field}
\langle E_i && (\mathbf{r}) E_j^\star (\mathbf{r}') \rangle = \langle E_i(\mathbf{r}) \rangle \langle E_j^\star (\mathbf{r}') \rangle \nonumber \\
&& + k_0^4 \int \langle G_{im}(\mathbf{r}-\mathbf{r}_1) \rangle \langle G_{jn}^\star(\mathbf{r}'-\mathbf{r}'_1) \rangle \Gamma_{mnrs}(\mathbf{r}_1,\mathbf{r}'_1,\mathbf{r}_2,\mathbf{r}'_2) \nonumber \\
  && \times \langle E_r(\mathbf{r}_2) E_s^\star (\mathbf{r}'_2) \rangle d\mathbf{r}_1 d\mathbf{r}'_1 d\mathbf{r}_2 d\mathbf{r}'_2 ,
\end{eqnarray}
where $\Gamma_{mnrs}(\mathbf{r}_1,\mathbf{r}'_1,\mathbf{r}_2,\mathbf{r}'_2)$ is the four-point irreducible vertex and the integral is taken over the volume occupied by the disordered medium. The first term in Eq.~(\ref{eqn:bethesalpeter-field}) corresponds to the ballistic (coherent) part of the propagating light, while the second (integral) term describes the multiple scattering process. For weak disorder ($k_0 \ell \gg 1$) and independent scattering (Eq.~(\ref{eq:disorder})), the so-called ladder approximation leads to
\begin{eqnarray}\label{eq:vertex}
\Gamma_{mnrs} && (\mathbf{r}_1,\mathbf{r}'_1,\mathbf{r}_2,\mathbf{r}'_2) \nonumber \\
 && \simeq \frac{6 \pi}{k_0^4 \ell} \delta(\mathbf{r}_1-\mathbf{r}'_1) \delta(\mathbf{r}_1-\mathbf{r}_2) \delta(\mathbf{r}'_1-\mathbf{r}'_2) \delta_{mr} \delta_{ns}.
\end{eqnarray}

As a source, we consider a point dipole at $\mathbf{r}_0$, defined as
\begin{equation}\label{eq:dipole_source}
j_k(\mathbf{r})=-i\omega p_k(\mathbf{r}) \delta(\mathbf{r}-\mathbf{r}_0) ,
\end{equation}
where $p_k(\mathbf{r})=\frac{1}{\mu_0 \omega^2}$ is the dipole moment oriented along direction $k$. This normalization was chosen such that the power radiated by a dipole in free space is 1 W. Making use of Eq.~(\ref{eq:field-green}), we finally arrive at a Bethe-Salpeter equation in terms of the Green function~\cite{Tatarski1961}
\begin{eqnarray}\label{eq:bethesalpeter-green}
\langle G_{ik} && (\mathbf{r}-\mathbf{r}_0) G_{jl}^\star (\mathbf{r}'-\mathbf{r}_0) \rangle = \langle G_{ik}(\mathbf{r}-\mathbf{r}_0) \rangle \langle G_{jl}^\star (\mathbf{r}'-\mathbf{r}_0) \rangle \nonumber \\
&& + \frac{6\pi}{\ell} \int \langle G_{im}(\mathbf{r}-\mathbf{r}_1) \rangle \langle G_{jn}^\star(\mathbf{r}'-\mathbf{r}_1) \rangle \nonumber \\
  && \times \langle G_{mk} (\mathbf{r}_1-\mathbf{r}_0) G_{nl}^\star (\mathbf{r}_1-\mathbf{r}_0) \rangle d\mathbf{r}_1 .
\end{eqnarray}

\begin{figure}
\begin{center}
	\includegraphics[width=0.45\textwidth]{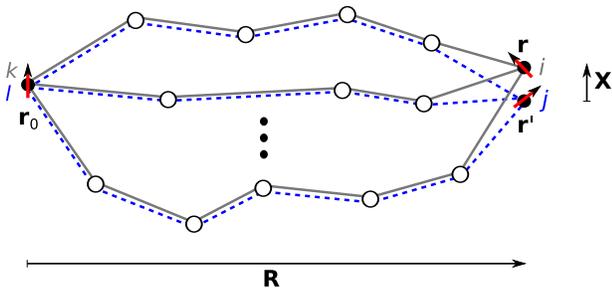}
\end{center}	
\caption{(Color online) Representation of the field correlation function for polarized waves in uncorrelated disordered media and notations used throughout the article. A dipole source is placed at point $\mathbf{r}_0$ and the field is observed at points $\mathbf{r}$ and $\mathbf{r}'$ (black dots). The solid gray and dashed blue lines represent the advanced and retarded averaged Green functions and the hollow dots represent scattering events. The input and output polarization states (represented by red arrows) are $(k,l)$ and $(i,j)$, respectively. The spatial field correlation function is derived within the ladder approximation (the advanced and retarded Green functions follow the same trajectories) and involves a sum over all possible scattering trajectories.
\label{fig1}}
\end{figure}

Eq.~(\ref{eq:bethesalpeter-green}) is represented schematically in Fig.~\ref{fig1}. In the ladder approximation, the polarization and coherence statistical properties of light is a result of a sum over all possible scattering trajectories.

%

\subsection{Resolution of the Bethe-Salpeter equation}

A key step for our study on the polarization and coherence of electromagnetic waves in disordered media is the resolution of Eq.~(\ref{eq:bethesalpeter-green}). The \textit{two-point} correlation matrix depends on the \textit{one-point} correlation matrix $\langle G_{mk} (\mathbf{r}_1-\mathbf{r}_0) G_{nl}^\star (\mathbf{r}_1-\mathbf{r}_0) \rangle$, which first needs to be calculated. By setting $\mathbf{r}'=\mathbf{r}$ in Eq.~(\ref{eq:bethesalpeter-green}), we obtain a self-consistent equation for $\langle G_{ik} (\mathbf{r}-\mathbf{r}_0) G_{jl}^\star (\mathbf{r}-\mathbf{r}_0) \rangle$ that can be solved and used afterwards to calculate the two-point correlation matrix. In reciprocal space, this equation reads
\begin{equation}\label{eqn:BS1}
D_{ijkl}(\mathbf{K})=S_{ijkl}(\mathbf{K})+\frac{6 \pi}{\ell} S_{ijmn}(\mathbf{K}) D_{mnkl}(\mathbf{K}),
\end{equation}
where $S_{ijkl}(\mathbf{K})=\int \langle G_{ik}(\mathbf{q}+\frac{\mathbf{K}}{2})\rangle \langle G_{jl}^\star(\mathbf{q}-\frac{\mathbf{K}}{2}) \rangle \frac{d\mathbf{q}}{(2\pi)^3}$ is the source term, and $D_{ijkl}(\mathbf{K})=\int \langle G_{ik}(\mathbf{q}+\frac{\mathbf{K}}{2}) G_{jl}^\star(\mathbf{q}-\frac{\mathbf{K}}{2}) \rangle \frac{d\mathbf{q}}{(2\pi)^3}$ is the diffuse term. Here, $\mathbf{q}$ and $\mathbf{K}$ are the Fourier transform variables of $\mathbf{X}=\mathbf{r}-\mathbf{r}'$ and $\mathbf{R}=(\mathbf{r}+\mathbf{r}')/2-\mathbf{r}_0$, respectively (see Fig.~\ref{fig1}).

The resolution of Eq.~(\ref{eqn:BS1}) requires us to find the nine eigenvalues $\lambda_p$ and eigenvectors $|kl\rangle_p$ of $\frac{6\pi}{\ell} S_{ijkl}$, such that we can write
\begin{equation}\label{eq:eigenSijkl}
\frac{6\pi}{\ell} S_{ijkl}(\mathbf{K})=\sum_{p=1}^9 \lambda_p |ij\rangle_p \langle kl|_p.
\end{equation}
Details on the resolution of the eigenvalue problem are given in Appendix~\ref{sec:A1}.

It is important to note that in previous works~\cite{Stephen1986, MacKintosh1988}, the authors have used a perturbation theory for \textit{non-degenerate} states~\cite{CohenTannoudji1991}. In the limit $K \rightarrow 0$, however, the eigenvalues have degenerary 1, 3 and 5, such that the approach employed is formally incorrect, as remarked also in Refs.~\cite{Ozrin1992, Muller2002}. This affects the analytical expressions obtained for the lineshape of the coherent backscattering cone, of main concern in these papers. The resulting correction may however be quantitatively small -- a fair agreement was actually found with previous experiments~\cite{Etemad1986} -- and it is likely that the main conclusions given on the respective contribution of polarized and depolarized light and the effect of Faraday rotation and natural optical activity on the cone lineshape remain valid. Similar considerations can be made on other previous results obtained using the same approach, for instance, in the context of diffusing wave spectroscopy~\cite{MacKintosh1989}. A reinvestigation of these phenomena, as partly done in Refs.~\cite{Ozrin1992, Amic1997}, is clearly not the scope of this paper. More importantly for us, the imprecise perturbation theory in Refs.~\cite{Stephen1986, MacKintosh1988} or more generally, the lack of complete expressions for the eigenvalues and eigenvectors of $\frac{6\pi}{\ell} S_{ijkl}(\mathbf{K})$ certainly excludes the possibility to derive analytically the field correlation matrix, which is the main purpose of the present paper.

We solve the eigenvalue problem \textit{exactly} to order $K^2$ from classical linear algebra, assuming that $K \ll q=|\mathbf{q}|$. This means that the distance between the two observation points should be much smaller than their distance to the source, which is verified in the cases considered in Sec.~\ref{sec:3}. We find six eigenvalues, three of them are doubly degenerate
\begin{eqnarray}\label{eq:eigenvalues}
&&\lambda_{1} \simeq  1-\frac{1}{3} K^2 \ell^2 , \quad  \lambda_{2} = \frac{1}{2}-\frac{3}{10} K^2 \ell^2, \nonumber \\ 
&&\lambda_{3,4} = \frac{1}{2}-\frac{1}{10} K^2 \ell^2 ,\quad  \lambda_{5,6} = \frac{7}{10}-\frac{23}{70} K^2 \ell^2, \nonumber \\
&&\lambda_{7,8} = \frac{7}{10}-\frac{13}{70} K^2 \ell^2 , \quad \lambda_{9} \simeq \frac{7}{10}-\frac{29}{210} K^2 \ell^2 ,
\end{eqnarray}
where $\lambda_1$ and $\lambda_9$ are given to order $K^2$. These eigenvalues differ from those given in Refs.~\cite{Stephen1986, MacKintosh1988} for the reason explained above. The corresponding nine eigenvectors, which form an orthonormal basis, have complicated expressions and are not given explicitly here. Nevertheless, one can show that in the limit $K \rightarrow 0$, they reduce to the following forms
\begin{eqnarray}\label{eq:eigenvectors}
&& |kl\rangle_{1} = \frac{1}{\sqrt{3}} \delta_{kl}, \nonumber \\
&& |kl\rangle_{2,3,4} = \frac{1}{\sqrt{2}} (\delta_{ka} \delta_{lb} - \delta_{kb} \delta_{la}), \nonumber \\
&& |kl\rangle_{5} = \frac{1}{\sqrt{2}} (\delta_{ka} \delta_{la} - \delta_{kb} \delta_{lb}), \nonumber \\
&& |kl\rangle_{6,7,8} = \frac{1}{\sqrt{2}} (\delta_{ka} \delta_{lb} + \delta_{kb} \delta_{la}), \nonumber \\
&& |kl\rangle_{9} = \frac{1}{\sqrt{6}} (\delta_{ka} \delta_{la} + \delta_{kb} \delta_{lb}- 2 \delta_{kc} \delta_{lc}),
\end{eqnarray}
where $a$, $b$ and $c$ are different from each other and take on the values 1, 2 and 3, which indicate the Cartesian coordinates. These eigenvectors are exactly those obtained by solving the eigenvalue problem for $K=0$ (see, e.g., Sec. 6.6.2 in Ref.~\cite{Akkermans2007}). As discussed further in Sec.~\ref{sec:3.1}, each of them describes a certain ``polarization eigenchannel'' that relates pairs of input field components ($k$ and $l$) with pairs of output field components ($i$ and $j$).

Using the \textit{complete} expressions for the eigenvalues and eigenvectors, the eigenvalue decomposition in Eq.~(\ref{eq:eigenSijkl}) allows us to rewrite Eq.~(\ref{eqn:BS1}) for each mode as
\begin{equation}\label{eq:Dm}
D_p=\frac{\ell}{6 \pi} \frac{\lambda_p}{1-\lambda_p},
\end{equation}
and to derive the one-point correlation matrix
\begin{equation}\label{eq:green_local_correlation_matrix}
D_{ijkl}(\mathbf{K})=\sum_{p=1}^9 D_p |ij\rangle_p \langle kl|_p.
\end{equation}
This concludes the first step of the derivation. The matrix $D_{ijkl}(\mathbf{K})$, valid up to the order $K^2$, describes the diffusion of the polarization (i.e. $\mathbf{r}'=\mathbf{r}$). We will come back on this specific aspect in Sec.~\ref{sec:3.1}.

The problem for the two-point correlation matrix $\langle G_{ik}(\mathbf{r}-\mathbf{r}_0) G_{jl}^\star (\mathbf{r}'-\mathbf{r}_0) \rangle$ can now be tackled. In reciprocal space, Eq.~(\ref{eq:bethesalpeter-green}) becomes
\begin{equation}\label{eqn:BS2}
Q_{ijkl}(\mathbf{q},\mathbf{K})=M_{ijkl}(\mathbf{q},\mathbf{K})+\frac{6 \pi}{\ell} M_{ijmn}(\mathbf{q},\mathbf{K}) D_{mnkl}(\mathbf{K})
\end{equation}
where $Q_{ijkl}(\mathbf{q},\mathbf{K})=\langle G_{ik}(\mathbf{q}+\frac{\mathbf{K}}{2}) G_{jl}^\star(\mathbf{q}-\frac{\mathbf{K}}{2}) \rangle$,  $M_{ijkl}(\mathbf{q},\mathbf{K})=\langle G_{ik}(\mathbf{q}+\frac{\mathbf{K}}{2}) \rangle \langle G_{jl}^\star(\mathbf{q}-\frac{\mathbf{K}}{2}) \rangle$, and $D_{ijkl}(\mathbf{K})$ was calculated above. Making use of Eq.~(\ref{eq:greentensor}) and assuming $K \ll q$, we can write
\begin{eqnarray}
Q_{ijkl}&&(\mathbf{q},\mathbf{K}) = \bigg[ (\delta_{ik}-\hat{q}_i \hat{q}_k) (\delta_{jl}-\hat{q}_j \hat{q}_l) \nonumber \\
&& + \frac{6\pi}{\ell} (\delta_{im}-\hat{q}_i \hat{q}_m) (\delta_{jn}-\hat{q}_j \hat{q}_n) D_{mnkl}(\mathbf{K}) \bigg] \nonumber \\
&& \times \langle G (\mathbf{q}+\frac{\mathbf{K}}{2}) \rangle \langle G^\star (\mathbf{q}-\frac{\mathbf{K}}{2}) \rangle.
\end{eqnarray}
The product of the averaged Green functions on the last line can be expanded in powers of $\mathbf{K}$, as shown in Appendix~\ref{sec:A1}, Eq.~(\ref{eq:greenproduct_Kexpansion}).

Finally, without loss of generality, we take the case of a dipole source oriented along the direction $k=l=1$ (meaning that it is the same dipole that supplies $G_{ik}$ and $G^\star_{jl}$, see Fig.~\ref{fig1}), and obtain an expression for the two-point field correlation matrix in reciprocal space, as
\begin{equation}
W_{ij}(\mathbf{q},\mathbf{K}) = \langle E_i (\mathbf{q}+\frac{\mathbf{K}}{2}) E_j^\star (\mathbf{q}-\frac{\mathbf{K}}{2}) \rangle = Q_{ij11}(\mathbf{q},\mathbf{K}).
\end{equation}

The complete expressions for the elements of $W_{ij}(\mathbf{q},\mathbf{K})$ are given in Appendix~\ref{sec:A2}. As such, the matrix contains all statistical information about the polarization and coherence of light generated by a dipole source in a disordered medium within the ladder and diffusion approximations. Due to the lengthy and complicated expressions, the inverse Fourier transform of $W_{ij}(\mathbf{q},\mathbf{K})$ cannot be easily handled analytically. It may nevertheless be used as a starting point in future theoretical studies. Analytical expressions can actually be obtained in certain limits. An important example is given in Sec.~\ref{sec:3.2}, where we investigate the spatial coherence of polarized light at large distances from the source.

\section{Results on polarization and coherence}\label{sec:3}

\subsection{Diffusion of polarization}\label{sec:3.1}

The eigenmode decomposition of $D_{ijkl}(\mathbf{K})$ in Eq.~(\ref{eq:green_local_correlation_matrix}) gives a clear physical understanding and a solid theoretical basis for the description of the diffusion of polarization (i.e. for $\mathbf{r}'=\mathbf{r}$) away from a polarized source. The eigensubspaces $|ij\rangle_p \langle kl|_p$ essentially describe the redistribution of the energy density from a given pair of input polarization components $k$ and $l$ onto pairs of output polarization components $i$ and $j$. In the limit $K \rightarrow 0$, i.e. far from the source (see Eq.~(\ref{eq:eigenvectors})), the polarization eigenchannels for $p=1$, $5$ and $9$ relate parallel components only (e.g., for $p=9$, $\langle 11|$ in input gives $|11 \rangle + |22 \rangle - |33 \rangle$ in output), while the remaining ones relate orthogonal components only (e.g., for $p=2$, $\langle 23|$ in input gives $|23 \rangle - |32 \rangle$ in output). In general, the redistribution of the polarization is much more complex, involving parallel and orthogonal field components and depending on the orientation of the source in a non-trivial way. This information is available in the complete expressions of the eigenvectors derived to order $K^2$ in the previous section.

Concurrently, the eigenvalues $D_p$ provide the characteristic length scales that describe the propagation of the energy density in the individual polarization eigenchannels. It is already known that polarization involves the existence of additional propagating modes~\cite{Stephen1986, Akkermans2007} and it was predicted that their respective energy density would decay exponentially on the scale of the collision time $\tau=\frac{\ell}{c}$, leaving only the scalar mode at long times. Here, we go beyond these results and provide analytical expressions for the spatial distribution of the energy density propagating through the individual polarization eigenchannels.

Writing the eigenvalues in Eq.~(\ref{eq:eigenvalues}) in the form $\lambda_p=A_p-B_p K^2 \ell^2$ and developing $\frac{1-\lambda_p(\mathbf{K})}{\lambda_p(\mathbf{K})}$ around $K=0$, Eq.~(\ref{eq:Dm}) can be rewritten as
\begin{equation}\label{diffuse_reciprocal}
\frac{6\pi}{\ell} \left( \frac{1}{A_p} -1 + \frac{B_p}{A_p^2} K^2 \ell^2 \right) D_p(\mathbf{K})=1.
\end{equation}
We define a quantity $U_p(\mathbf{K})=\frac{6\pi}{c} D_p(\mathbf{K})$ that has the dimension of an energy density ($\text{J}.\text{m}^{-3}$). After inverse Fourier transform of Eq.~(\ref{diffuse_reciprocal}), we find that the energy density associated to each polarization eigenchannel obeys a classical diffusion equation with an effective attenuation term, $-\mathcal{D}_p \nabla^2 U_p(\mathbf{R}) + \mu_{a,p} c U_p(\mathbf{R}) = \delta(\mathbf{R})$, where $\mathcal{D}_p= \frac{B_p}{A_p^2} c \ell$ and $\mu_{a,p}=\frac{1}{\ell} \left( \frac{1}{A_p} -1 \right)$ are the diffusion constant and the attenuation coefficient of the $p$th polarization eigenchannel. The solution of the diffusion equation is
\begin{equation}\label{eq:diffusionsolution}
U_p(\mathbf{R}) = \frac{1}{4\pi \mathcal{D}_p R} \exp \left[ - \frac{R}{\ell_{\text{eff},p}} \right],
\end{equation}
with $R=|\mathbf{R}|$ and $\ell_{\text{eff},p}=\sqrt{\frac{\mathcal{D}_p}{\mu_{a,p} c}}$ an effective attenuation length.

Table~\ref{tab:diffusion_pola} summarizes the diffusion constants and effective attenuation lengths for the different polarization eigenchannels. As expected, we recover the expression for the diffusion of the classical intensity through the eigenchannel $p=1$, that corresponds to the scalar mode, which stems from the requirement that energy should be conserved throughout the multiple scattering process~\cite{Akkermans2007}. It is the only remaining mode at long distances from the source and therefore justifies the use of scalar theories for the propagation of the average energy density in disordered media. The diffusion through the other polarization eigenchannels $p=2...9$ exhibit varying diffusivities and are all attenuated on a length scale of the order of the mean free path ($\ell_{\text{eff},p} \approx \ell$). This attenuation translates the progressive depolarization of light away from a polarized source. It is worth emphasizing that these results apply to \textit{uncorrelated} disordered media, for which the transport mean free path $\ell_t$, that is the relevant length scale for diffusive processes, is exactly equal to the scattering mean free path $\ell$. For correlated disordered media (or systems composed of finite-size scatterers), we expect the effective attenuation coefficients $\ell_{\text{eff},p}$ and diffusion constants $ \mathcal{D}_p$ to depend of $\ell_t$, possibly in a non-trivial way. We leave this investigation for future research.

\begin{table}
\centering
\begin{tabular}{c|c|c|c|c|c|c}
$p$ & 1 & 2 & 3,4 & 5,6 & 7,8 & 9 \\[0.2cm] \hline
$\mathcal{D}_p$ & $\frac{1}{3}c\ell$ & $\frac{6}{5}c\ell$ & $\frac{2}{5}c\ell$ & $\frac{230}{343}c\ell$ & $\frac{130}{343}c\ell$ & $\frac{290}{1029}c\ell$ \\[0.2cm] \hline
$\ell_{\text{eff},p}$ & $\infty$ & $\sqrt{\frac{6}{5}} \ell$ & $\sqrt{\frac{2}{5}} \ell$ & $\sqrt{\frac{230}{147}} \ell$ & $\sqrt{\frac{130}{147}} \ell$ & $\sqrt{\frac{290}{441}} \ell$
\end{tabular}
\caption{Summary of the diffusion constants $\mathcal{D}_p$ and effective attenuation lengths $\ell_{\text{eff},p}$ characterizing the diffusion properties of the energy density through the individual polarization eigenchannels, Eq.~(\ref{eq:diffusionsolution}). The eigenchannel $p=1$ corresponds to the scalar (Goldstone) mode. It is the only remaining mode at large distances from the source, the other ones being exponentially attenuated on a length scale of the order of the mean free path.}
\label{tab:diffusion_pola}
\end{table}

In sum, we have calculated the eigenmodes that govern the diffusion of the polarization and the characteristic length scales describing the energy propagation through the individual polarization eigenchannels. This provides an important insight into the way light is depolarized away from a dipole source and constitutes the first important result of this article.

\subsection{Coherence of polarized light in disordered media}\label{sec:3.2}

Another fundamental property of polarized light is spatial coherence, which, as explained previously, describes the correlation between parallel field components at two different points of space. Although the complete expressions for the two-point field correlation functions $W_{ij}(\mathbf{q},\mathbf{K})$ given in Appendix~\ref{sec:A2} are rather complicated, it is still possible to investigate analytically the spatial light coherence at large distances from the source.

Keeping only the leading term to order $K^2$, one finds a simple relation for the field correlation in reciprocal space
\begin{equation}\label{eq:fieldcorrelationbulk}
W_{ij}(\mathbf{q},\mathbf{K}) = \frac{(\delta_{ij}-\hat{q}_i \hat{q}_j)}{K^2 \ell^2}  \langle G(\mathbf{q}) \rangle \langle G^\star(\mathbf{q}) \rangle.
\end{equation}
As shown in Appendix~\ref{sec:A3}, this expression correctly leads to the classical result for the diffusion of the energy density in real space. It follows from Eq.~(\ref{eq:fieldcorrelationbulk}) that, at large distances from the source, the two-point correlation function is independent of the orientation of the source dipole. Let us stress that this is not the case concerning the other terms in the two-point field correlation matrix that contribute at shorter distances, indicating that information about the dipole source orientation is still contained in the two-point field correlation matrix. This aspect could be investigated in future studies.

The two-point field correlation matrix in real space, $W_{ij}(\mathbf{R},\mathbf{X})= \langle E_i(\mathbf{R}+\frac{\mathbf{X}}{2}) E_j^\star(\mathbf{R}-\frac{\mathbf{X}}{2}) \rangle$, can be obtained by inverse Fourier transform of Eq.~(\ref{eq:fieldcorrelationbulk}). Details are given in the Appendix~\ref{sec:A4}. Without loss of generality, we assume that the two observation points are oriented along direction $1$, $\mathbf{X}=X\hat{u}_1$, and find
\begin{equation}\label{eq:coherence}
W_{ij}(\mathbf{R},\mathbf{X})=
\begin{cases}
\frac{f^{(0)}(\mathbf{X})}{8 \pi^2 \ell R} \quad & (i=j=1) \\
\frac{f^{(1)}(\mathbf{X})-f^{(0)}(\mathbf{X})}{16 \pi^2 \ell R} \quad &(i=j=2,3) \\
 0 \quad &(i \neq j),
\end{cases}
\end{equation}
where
\begin{eqnarray}\label{eq:f0}
f^{(0)}(\mathbf{X}) &=& \frac{1}{2 (k_0 X)^3 (k_0 \ell)} \nonumber \\
&& \times \Big[ 2 - k_0 \ell n_e^2 \left( k_0 X n_e^\star -i \right) \exp \left[ -ik_0 X n_e^\star \right] \nonumber \\
&& - k_0 \ell n_e^{\star2} \left( k_0 X n_e +i \right) \exp \left[ ik_0 X n_e \right] \Big]
\end{eqnarray}
and
\begin{equation}\label{eq:f1}
f^{(1)}(\mathbf{X})=\frac{1}{2 i k_0 X} \Big( \exp[ik_0 X n_e] - \exp[-ik_0 X n_e^\star] \Big),
\end{equation}
where the effective refractive index is $n_e=1+i/(2k_0\ell)$, the final expressions~(\ref{eq:f0}) and (\ref{eq:f1}) being valid up to the order $(k_0\ell)^{-1}$.

\begin{figure}
\begin{center}
	\includegraphics[width=0.45\textwidth]{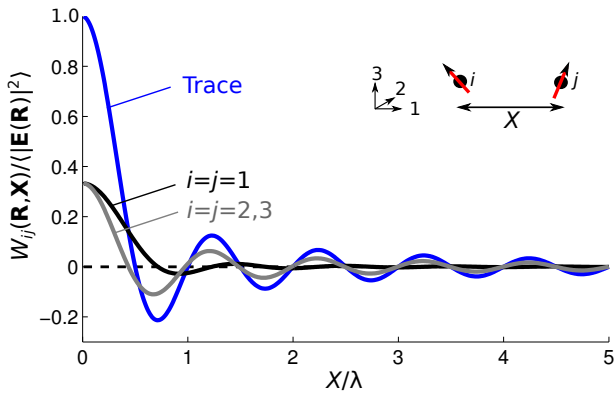}
\end{center}	
\caption{(Color online) Normalized two-point field correlation in an uncorrelated disordered medium. The free-space wavevector is $k_0=\frac{2\pi}{\lambda}$ and the mean free path is $\ell=20 \lambda$ (e.g. for $\lambda=532$ nm, $\ell \approx 10.6$ $\mu$m). The $\mathbf{X}$ vector is oriented along the direction $1$. The correlation between the parallel field components oriented along direction $1$ differs from that for parallel components oriented along directions $2$ and $3$. The correlation between orthogonal components is zero. The trace of the correlation matrix yields the well-known expression for the correlation function of scalar waves, Eq.~(\ref{eq:classical_coherence}).
\label{fig2}}
\end{figure}

A first observation on the field correlation matrix is that the non-diagonal terms vanish, signifying the absence of correlation between orthogonal field components in two points. This is a natural consequence of the light depolarization investigated in the previous section, where the correlation between orthogonal field components should disappear at large distances from the source. A second observation is that the correlation for parallel field components strongly depends on their orientation with respect to $\mathbf{X}$ (here along direction $1$). The correlation functions, normalized to $\langle |\mathbf{E}(\mathbf{R})|^2 \rangle=\frac{1}{8 \pi^2 \ell R}$, are plotted in Fig.~\ref{fig2}. The field correlation function of the components parallel to $\mathbf{X}$ is broader at short distances and is attenuated faster than the correlation for components normal to $\mathbf{X}$. This fundamental property of electromagnetic fields in disordered media is the second important result of this paper.

Finally, we calculate the trace of $W_{ij} (\mathbf{R},\mathbf{X})$, which, from Eq.~(\ref{eq:coherence}), simplifies to $\frac{f^{(0)}(\mathbf{X})}{8 \pi^2 \ell R}$, leading to the same expression as that known for the correlation function of scalar waves~\cite{Shapiro1986, Sheng2010}
\begin{equation}\label{eq:classical_coherence}
\text{Tr} \left[ W_{ij} (\mathbf{R},\mathbf{X}) \right] = \langle | \mathbf{E}(\mathbf{R}) |^2 \rangle \text{sinc}\left[k_0 X \right] \exp \left[-\frac{X}{2\ell} \right].
\end{equation}
This is expected as the degree of coherence of the field produced by statistically stationary, homogeneous, and isotropic current distributions was shown to be universal~\cite{Setala2003}.

\section{Conclusion}\label{sec:4}

To conclude, we have investigated theoretically the polarization and spatial coherence properties of electromagnetic waves produced by a dipole source in an uncorrelated disordered medium, using a multiple scattering theory for polarized light within the ladder and diffusion approximations. Our calculations have been performed to orders $K^2$ (i.e. at large distances from the source) and $(k_0\ell)^{-1}$ (i.e. for weak disorder), and assuming that the distance between the observation points is much smaller than their distance to the source ($X \ll R$).

Our first result is the explicit calculation of the eigenmodes that govern the diffusion of the polarization and the finding that the energy density propagation through the individual polarization eigenchannels can be described by a diffusion equation with an attenuation term, see Eq.~(\ref{eq:diffusionsolution}) and Table~\ref{tab:diffusion_pola}. The exponential attenuation, found to be on the scale of a mean free path, measures the progressive light depolarization away from the source. This eigenmode decomposition gives a solid theoretical basis for the description of the transport of the energy density for polarized light in disordered media. It could be particularly relevant for the optical characterization of complex materials, where polarization-resolved measurements in reflection geometries could bring further information into, for instance, the presence of large-scale heterogeneities~\cite{Tuchin2006}.

Our second result is the derivation of new analytical formulas for the spatial coherence of light in multiply-scattering media, see Eq.~(\ref{eq:coherence}). Interestingly, the correlation function between field components was found to depend strongly on their orientation with respect to the two observation points. The classical expression of the correlation function for scalar waves is recovered by summing over all directions. We have therefore evidenced a very fundamental property of polarized waves in disordered media. This, in turn, strongly emphasizes the need in taking polarization effects into account when mesoscopic scales are part of the physics. Our results have no bearing whatsoever on the problem recently raised on the three-dimensional Anderson localization of light~\cite{Skipetrov2013}, since one should go beyond the ladder approximation, but they suggest that many optical phenomena in disordered media simply cannot be captured by scalar theories.

We believe that the theoretical results obtained here could be a starting point for further studies on polarization effects in complex systems and find interest in optical imaging and spectroscopy techniques for medical and material science applications. On this respect, several aspects may deserve particular attention.

First, we have seen that the theoretical description of the multiple scattering of polarized light in general terms implies lengthy and complicated expressions that cannot be easily handled analytically. An interesting follow-up of our study could therefore be to calculate \textit{numerically} the spatial field correlation matrix in real space, as to gain insight into the polarization and coherence properties of light in disordered media (e.g., at shorter distances from the source) and validate the theoretical predictions reported here. This could be obtained by performing either a numerical inverse Fourier transform of the spatial field correlation matrix $W_{ij}(\mathbf{q},\mathbf{K})$ given in Appendix~\ref{sec:A2} or numerical calculations of realistic systems using the coupled dipole method~\cite{Lax1952}.

Second, it is important to note that the derivation here was made for \textit{uncorrelated} disordered media. Systems possessing a structural correlation or composed of large particles are expected to change the light diffusion characteristics. For scalar waves, one can introduce a new length scale, the transport mean free path $\ell_t$, that is generally longer than the scattering mean free path $\ell$~\cite{vanRossum1999, Akkermans2007}. However, it is not obvious how our results on the diffusion of the polarization (Table~\ref{tab:diffusion_pola}), for instance, would be modified by the presence of structural correlations (apart from replacing $\ell$ by $\ell_t$). This aspect is intimately related to the recent interest in understanding how the specific morphology of disordered media affect their optical properties at mesoscopic scales~\cite{Haefner2010, Carminati2010}.

Finally, as mentioned earlier, the notions of polarization and coherence have been essentially used in the fields of beam optics and atmopheric turbulence, that is, for paraxial light propagation~\cite{Andrews2005}. The well-known degrees of polarization and of coherence have been introduced as to provide one-quantity measures of the properties of fluctuating two-dimensional fields. For three-dimensional fields, since the field correlation matrix relates two fields only, their definition is more subtle~\cite{Setala2002, Tervo2003, Ellis2005, Refregier2012, Aunon2013, Dogariu2013b}. We hope that our work will motivate further investigations along these lines.

\section{Acknowledgements}

This work is supported by LABEX WIFI (Laboratory of Excellence within the French Program "Investments for the Future") under references ANR-10-LABX-24 and ANR-10-IDEX-0001-02 PSL$^\star$.

\appendix

\section{Eigenvalue decomposition of $S_{ijkl}(\mathbf{K})$}\label{sec:A1}

In this Appendix, we explain in detail the eigenvalue decomposition of $S_{ijkl}(\mathbf{K})$ in Eq.~(\ref{eqn:BS1}) to order $K^2$. The first part of the derivation is similar to that presented in Refs.~\cite{Stephen1986, MacKintosh1988} while the second part differs significantly.

We begin with
\begin{equation}
S_{ijkl}(\mathbf{K})=\int \langle G_{ik}(\mathbf{q}+\frac{\mathbf{K}}{2})\rangle \langle G_{jl}^\star(\mathbf{q}-\frac{\mathbf{K}}{2}) \rangle \frac{d\mathbf{q}}{(2\pi)^3}.
\end{equation}
Separating the polarized and scalar parts of the dyadic Green function as in Eq.~(\ref{eq:greentensor}) and the integral over $\mathbf{q}$ as an integral over $q=|\mathbf{q}|$ and an angular average over $\hat{\mathbf{q}}=\frac{\mathbf{q}}{q}$, we can write
\begin{eqnarray}
S_{ijkl}&&(\mathbf{K})=\frac{4\pi}{(2\pi)^3} \langle (\delta_{ik}-\hat{q}_i \hat{q}_k) (\delta_{jl}-\hat{q}_j \hat{q}_l) \nonumber \\
&& \times \int_0^\infty \langle G(\mathbf{q}+\frac{\mathbf{K}}{2})\rangle \langle G^\star(\mathbf{q}-\frac{\mathbf{K}}{2}) \rangle q^2 dq \rangle_{\hat{\mathbf{q}}}.
\end{eqnarray}
Here, the $\mathbf{K}$-dependence of the polarization term has been neglected under the assumption that $K \ll q$.

The average Green functions can be developed in powers of $\mathbf{K}$ around $\mathbf{0}$, yielding, up to order $K^2$
\begin{eqnarray}\label{eq:greenproduct_Kexpansion}
\langle G && (\mathbf{q}+\frac{\mathbf{K}}{2}) \rangle \langle G^\star (\mathbf{q}-\frac{\mathbf{K}}{2}) \rangle = \langle G (\mathbf{q}) \rangle \langle G^\star (\mathbf{q}) \rangle \nonumber \\
&& \times \bigg( 1 - (\mathbf{q} \cdot \mathbf{K}) \big( \langle G (\mathbf{q}) \rangle - \langle G^\star (\mathbf{q}) \rangle \big) \nonumber \\
&& + (\mathbf{q} \cdot \mathbf{K})^2 \big( \langle G (\mathbf{q}) \rangle^2 + \langle G^\star (\mathbf{q}) \rangle^2 - \langle G (\mathbf{q}) \rangle \langle G^\star (\mathbf{q}) \rangle \big) \nonumber \\
&& + \frac{K^2}{4} \big( \langle G (\mathbf{q}) \rangle + \langle G^\star (\mathbf{q}) \rangle \big) \bigg).
\end{eqnarray}
The integral over $q$ can then be solved using the residue theorem
\begin{eqnarray}\label{eq:residue}
\frac{4\pi}{(2\pi)^3} && \int_0^\infty \langle G(\mathbf{q}+\frac{\mathbf{K}}{2})\rangle \langle G^\star(\mathbf{q}-\frac{\mathbf{K}}{2}) \rangle q^2 dq = \nonumber \\
&& \frac{\ell}{6\pi} \frac{3}{2} \bigg( 1 - i (\hat{\mathbf{q}} \cdot \mathbf{K}) \ell - (\hat{\mathbf{q}} \cdot \mathbf{K})^2\ell^2 + \frac{K^2}{8 k_0^2} \bigg).
\end{eqnarray}
Having $K \ll k_0$ (i.e. $R \gg \lambda$), the last term can be neglected, yielding
\begin{eqnarray}\label{eq:Sijkl_1}
S_{ijkl}&&(\mathbf{K}) = \frac{\ell}{6\pi} \frac{3}{2} \langle (\delta_{ik}-\hat{q}_i \hat{q}_k) (\delta_{jl}-\hat{q}_j \hat{q}_l) \nonumber \\
&& (1 - i \hat{q}_r K_r \ell -\hat{q}_m \hat{q}_n K_m K_n \ell^2) \rangle_{\hat{\mathbf{q}}}
\end{eqnarray}
Expanding Eq.~(\ref{eq:Sijkl_1}) and noting that $\langle \hat{q}_i^2 \rangle_{\hat{\mathbf{q}}}=\frac{1}{3}$, $\langle \hat{q}_i^4 \rangle_{\hat{\mathbf{q}}}=\frac{1}{5}$, $\langle \hat{q}_i^6 \rangle_{\hat{\mathbf{q}}}=\frac{1}{7}$, $\langle \hat{q}_i^2 \hat{q}_j^2 \rangle_{\hat{\mathbf{q}}}=\frac{1}{15}$, $\langle \hat{q}_i^2 \hat{q}_j^4 \rangle_{\hat{\mathbf{q}}}=\frac{1}{35}$ and $\langle \hat{q}_i^2 \hat{q}_j^2 \hat{q}_k^2 \rangle_{\hat{\mathbf{q}}}=\frac{1}{105}$ for $ i \neq j \neq k$, and angular averages containing odd powers of $\hat{q}_i$ equal zero, we reach this final expression
\begin{eqnarray}
S_{ijkl}&&(\mathbf{K}) = \frac{\ell}{6\pi} \frac{3}{2} \bigg( \frac{1}{3} \delta_{ik} \delta_{jl} + \frac{1}{15} \left( \delta_{ij}\delta_{kl} + \delta_{ik}\delta_{jl} + \delta_{il}\delta_{jk} \right) \nonumber \\
&& - \frac{1}{5} \delta_{ik} \delta_{jl} K^2 \ell^2 + \frac{2}{15} \delta_{ik} K_j K_l \ell^2 + \frac{2}{15} \delta_{jl} K_i K_k \ell^2 \nonumber \\
&& -  \frac{1}{105} \left( \delta_{ij} \delta_{kl} + \delta_{ik}\delta_{jl} + \delta_{il}\delta_{jk} \right) K^2 \ell^2 \nonumber \\
&& - \frac{2}{105} ( \delta_{ij} K_k K_l + \delta_{ik} K_j K_l + \delta_{il} K_j K_k \nonumber \\
&& + \delta_{kl} K_i K_j + \delta_{jk} K_i K_l + \delta_{jl} K_i K_k ) \ell^2 \bigg).
\end{eqnarray}

We now perform the eigenvalue decomposition
\begin{equation}\label{eq:eigendecomposition}
\frac{6\pi}{\ell} S_{ijkl}(\mathbf{K}) = \sum_{p=1}^9 \lambda_p |ij \rangle_p \langle kl|_p,
\end{equation}
where $\lambda_p$ and $ |kl \rangle_p$ are the $p$th eigenvalue and eigenvector of $\frac{6\pi}{\ell} S_{ijkl}(\mathbf{K})$, respectively. The eigenvalues can be found easily from the characteristic polynomial of the matrix. The results are given in Eq.~(\ref{eq:eigenvalues}), with $\lambda_{1,9}=\frac{1}{140} (119 - 33 K^2 \ell^2 \mp \sqrt{441+K^2 \ell^2 (-574+361 K^2 \ell^2)})$.

The calculation of the eigenvectors is much more involved due to the complexity of the derived expressions, in particular for $p=1$ and $9$. The problem can be tackled by solving the eigenvalue problem in each eigensubspace independently first \textit{excluding} $p=1$ and $9$,
\begin{equation}
\left( \frac{6 \pi}{\ell} S_{ijkl} - \lambda_p \delta_{ik} \delta_{jl} \right) | kl \rangle_p = 0,
\end{equation}
and imposing that all eigenvectors (including those of the degenerate eigenvalues) should be orthogonal to each other
\begin{equation}
\langle kl |_p |kl \rangle_q = \delta_{pq}.
\end{equation}
Then, the problem for the eigenvectors $p=1$ and $9$ can be simplified by imposing them to be orthogonal to all other eigenvectors and finally by solving the eigenvalue problem on this restricted set of solutions. A complete set of orthonormal eigenvectors (not shown here), which verify Eq.~(\ref{eq:eigendecomposition}), is obtained.

\section{Two-point field correlation matrix in reciprocal space}\label{sec:A2}

In this Appendix, we provide the complete expressions obtained for the two-point field correlation matrix, $W_{ij}(\mathbf{q},\mathbf{K})=\langle E_i(\mathbf{q}+\frac{\mathbf{K}}{2}) E_j^\star(\mathbf{q}-\frac{\mathbf{K}}{2}) \rangle $. We find that it can be written in the form
\begin{eqnarray}\label{eq:two-point_field_correlation_complete_1}
W_{ij}(\mathbf{q},\mathbf{K})  &&= \bigg( \frac{\mathcal{C}_{ij}^{(0)}(\hat{\mathbf{q}})}{K^2 \ell^2}  + \frac{7 \mathcal{C}_{ij}^{(1)}(\hat{\mathbf{q}},\mathbf{K}) }{K^4(35+13 K^2 \ell^2)}  \nonumber \\
&&+ \frac{70 \mathcal{C}_{ij}^{(2)}(\hat{\mathbf{q}},\mathbf{K})}{K^4(21+13 K^2 \ell^2)}  + \frac{35\mathcal{C}_{ij}^{(3)}(\hat{\mathbf{q}},\mathbf{K}) }{K^4(21+23K^2 \ell^2)} \bigg) \nonumber \\
&& \times \langle G(\mathbf{q}+\frac{\mathbf{K}}{2}) \rangle \langle G^\star(\mathbf{q}-\frac{\mathbf{K}}{2}) \rangle,
\end{eqnarray}
when $i=j$, and
\begin{eqnarray}\label{eq:two-point_field_correlation_complete_2}
W_{ij}&& (\mathbf{q},\mathbf{K}) = \nonumber \\
&& \frac{5 \mathcal{C}_{ij}^{(4)}(\hat{\mathbf{q}},\mathbf{K})}{K^2 \ell^2 (35+13 K^2 \ell^2) (21+13 K^2 \ell^2) (21+23K^2 \ell^2)} \nonumber \\
&& \times \langle G(\mathbf{q}+\frac{\mathbf{K}}{2}) \rangle \langle G^\star(\mathbf{q}-\frac{\mathbf{K}}{2}) \rangle,
\end{eqnarray}
when $i \neq j$. The average Green functions can be developed to order $K^2$ as in Eq.~(\ref{eq:greenproduct_Kexpansion}). In Eqs.~(\ref{eq:two-point_field_correlation_complete_1}-\ref{eq:two-point_field_correlation_complete_2}), $\mathcal{C}_{ij}^{(n)}(\mathbf{K},\mathbf{q})$ are prefactor functions, given by
\begin{equation}
\mathcal{C}_{ij}^{(0)}(\hat{\mathbf{q}}) =\delta_{ij} - \hat{q}_i \hat{q}_j,
\end{equation}
for all $i$ and $j$, and
\begin{widetext}
\begin{eqnarray}
& \mathcal{C}_{11}^{(1)}(\hat{\mathbf{q}},\mathbf{K}) =& (4 K_1^2 - K_2^2 - K_3^2) \big( 10 K_2 K_3 \hat{q}_1^2 \hat{q}_2 \hat{q}_3 + 10 K_1 \hat{q}_1 (-1 + \hat{q}_1^2) (K_2 \hat{q}_2 + K_3 \hat{q}_3) - K_2^2 (1-(1+5\hat{q}_2^2) \hat{q}_1^2) \nonumber \\
&& - K_3^2 (1-(1+5 \hat{q}_3^2) \hat{q}_1^2) + K_1^2 (4-\hat{q}_1^2(4+5(1-\hat{q}_1^2))) \big), \\
& \mathcal{C}_{11}^{(2)}(\hat{\mathbf{q}},\mathbf{K}) =& 2 K_1 \big( K_1 (-1+\hat{q}_1^2)+ \hat{q}_1(K_2 \hat{q}_2 + K_3 \hat{q}_3) \big) \big( K_1 \hat{q}_1 (K_2 \hat{q}_2 + K_3 \hat{q}_3) + (K_2^2 +K_3^2) (1-\hat{q}_1^2) \big), \\
& \mathcal{C}_{11}^{(3)}(\hat{\mathbf{q}},\mathbf{K}) =& K_3^4 + 2 K_2^3 \hat{q}_1 \hat{q}_2 (K_1-K_1 \hat{q}_1^2 + K_3 \hat{q}_1 \hat{q}_3) + K_2^4 (1+ \hat{q}_1^4 - \hat{q}_1^2 (2 + \hat{q}_3^2)) + 2 K_2 K_3 \hat{q}_1 \hat{q}_2 \nonumber \\
&& \times ( 2 K_1^2 \hat{q}_1 \hat{q}_3 + K_3^2 \hat{q}_1 \hat{q}_3 + K_1 K_3 (1-\hat{q}_1^2) )  - K_3^2 \hat{q}_1 (K_1^2 \hat{q}_1 (\hat{q}_2^2 - \hat{q}_3^2) - 2 K_1 K_3 \hat{q}_3 (1-\hat{q}_1^2) + K_3^2 \hat{q}_1 (1+ 2 \hat{q}_2^2 + \hat{q}_3^2)) \nonumber \\
&& + K_2^2 (K_1^2 \hat{q}_1^2 (\hat{q}_2^2 - \hat{q}_3^2) + 2 K_1 K_3 \hat{q}_1 \hat{q}_3 (1-\hat{q}_1^2) + K_3^2 (1-\hat{q}_1^2)(-1+3(1-\hat{q}_1^2))),
\end{eqnarray}

\begin{eqnarray}
& \mathcal{C}_{22}^{(1)}(\hat{\mathbf{q}},\mathbf{K}) =& (4 K_1^2 - K_2^2 - K_3^2) \big( (1-\hat{q}_2^2) (K_3^2 + 10 K_1 K_2 \hat{q}_1 \hat{q}_2 - K_2^2 (4-5 \hat{q}_2^2) + K_1^2 (1+5 \hat{q}_2^2) ) \nonumber \\
&& - 10 K_3 \hat{q}_2 \hat{q}_3 (K_1 \hat{q}_1 \hat{q}_2 - K_2 (1-\hat{q}_2^2)) + 5 (K_1^2 - K_3^2) \hat{q}_2^2 \hat{q}_3^2 \big), \\
& \mathcal{C}_{22}^{(2)}(\hat{\mathbf{q}},\mathbf{K}) =& 2 K_1 (K_1 \hat{q}_1 \hat{q}_2 - K_2 (1-\hat{q}_2^2) + K_3 \hat{q}_2 \hat{q}_3) ((K_2^2+K_3^2) \hat{q}_1 \hat{q}_2+ K_1 (K_2 (1-\hat{q}_2^2) - K_3 \hat{q}_2 \hat{q}_3)), \\
& \mathcal{C}_{22}^{(3)}(\hat{\mathbf{q}},\mathbf{K}) =& 2 K_1 (K_2^2 + K_3^2) \hat{q}_1 \hat{q}_2 (K_2 (1-\hat{q}_2^2) - K_3 \hat{q}_2 \hat{q}_3) + K_1^2 ((K_2^2 - K_3^2)(1-\hat{q}_2^2)^2 - 4 K_2 K_3 \hat{q}_2 \hat{q}_3 (1-\hat{q}_2^2) \nonumber \\
&& -(K_2^2-K_3^2) \hat{q}_2^2 \hat{q}_3^2) + (K_2^2 +K_3^2)  (K_3^2 (-1+(2+\hat{q}_1^2) \hat{q}_2^2- \hat{q}_2^4) - 2 K_2 K_3 \hat{q}_2 \hat{q}_3 (1-\hat{q}_2^2) + K_2^2 \hat{q}_2^2 (\hat{q}_1^2 - \hat{q}_3)^2 ), \nonumber \\
\end{eqnarray}

\begin{eqnarray}
& \mathcal{C}_{12}^{(4)}(\hat{\mathbf{q}},\mathbf{K}) =& (-3087 \hat{q}_1 \hat{q}_2 + 147 \ell^2 (14 K_1 (K_2 (1 - \hat{q}_2^2 + \hat{q}_1^2 (-1 + 2 \hat{q}_2^2)) + K_3 (-1 + 2 \hat{q}_1^2) \hat{q}_2 \hat{q}_3) \nonumber \\
&& - 3 K_1^2 \hat{q}_1 \hat{q}_2 (9 + 28 \hat{q}_2^2 + 28 \hat{q}_3^2) - \hat{q}_1 (14 K_2 K_3 (1 - 2 \hat{q}_2^2) \hat{q}_3 + K_3^2 \hat{q}_2 (13 + 70 \hat{q}_2^2 + 56 \hat{q}_3^2) \nonumber \\
&& + K_2^2 \hat{q}_2 (27 + 56 \hat{q}_2^2 + 70 \hat{q}_3^2))) + 13 \ell^6 K^2 (322 K_1^3 (K_2 (1 - \hat{q}_1^2 + (-1 + 2 \hat{q}_1^2) \hat{q}_2^2) + K_3 (-1 + 2 \hat{q}_1^2) \hat{q}_2 \hat{q}_3) \nonumber \\
&& + 70 K_1 (K_2^2 + K_3^2) (K_2 (1 - \hat{q}_1^2 + (-1 + 2 \hat{q}_1^2) \hat{q}_2^2) + K_3 (-1 + 2 \hat{q}_1^2) \hat{q}_2 \hat{q}_3) - 23 K_1^4 \hat{q}_1 \hat{q}_2 (-3 + 28 \hat{q}_2^2 + 28 \hat{q}_3^2) \nonumber \\
&& - (K_2^2 + K_3^2) \hat{q}_1 (70 K_2 K_3 (-1 + 2 \hat{q}_2^2) \hat{q}_3 + K_2^2 \hat{q}_2 (-69 + 252 \hat{q}_2^2 + 182 \hat{q}_3^2) + K_3^2 \hat{q}_2 (1 + 182 \hat{q}_2^2 + 252 \hat{q}_3^2)) \nonumber \\
&& - 2 K_1^2 \hat{q}_1 (91 K_2 K_3 (1 - 2 \hat{q}_2^2) \hat{q}_3 + K_3^2 \hat{q}_2 (-34 + 287 \hat{q}_2^2 + 196 \hat{q}_3^2) + K_2^2 \hat{q}_2 (57 + 196 \hat{q}_2^2 + 287 \hat{q}_3^2))) \nonumber \\
&& + 7 \ell^4 (868 K_1^3 (K_2 (1 - \hat{q}_1^2 + (-1 + 2 \hat{q}_1^2) \hat{q}_2^2) + K_3 (-1 + 2 \hat{q}_1^2) \hat{q}_2 \hat{q}_3) + 672 K_1 (K_2^2 + K_3^2) \nonumber \\
&& \times (K_2 (1 - \hat{q}_1^2 + (-1 + 2 \hat{q}_1^2) \hat{q}_2^2) + K_3 (-1 + 2 \hat{q}_1^2) \hat{q}_2 \hat{q}_3) + K_1^4 \hat{q}_1 \hat{q}_2 (25 - 3024 \hat{q}_2^2 - 3024 \hat{q}_3^2) \nonumber \\
&& - (K_2^2 + K_3^2) \hat{q}_1 (28 K_2 K_3 (-1 + 2 \hat{q}_2^2) \hat{q}_3 + K_2^2 \hat{q}_2 (-25 + 1484 \hat{q}_2^2 + 1456 \hat{q}_3^2) + K_3^2 \hat{q}_2 (3 + 1456 \hat{q}_2^2 + 1484 \hat{q}_3^2)) \nonumber \\
&& - 2 K_1^2 \hat{q}_1 (84 K_2 K_3 (1 - 2 \hat{q}_2^2) \hat{q}_3 + K_3^2 \hat{q}_2 (-11 + 2142 \hat{q}_2^2 + 2058 \hat{q}_3^2) + K_2^2 \hat{q}_2 (73 + 2058 \hat{q}_2^2 + 2142 \hat{q}_3^2)))),
\end{eqnarray}

\begin{eqnarray}
& \mathcal{C}_{23}^{(4)}(\hat{\mathbf{q}},\mathbf{K}) =& -3087 \hat{q}_2 \hat{q}_3 + 147 \ell^2 (K_2^2 \hat{q}_2 \hat{q}_3 (43 - 56 \hat{q}_2^2 - 70 \hat{q}_3^2) + 14 K_2 (K_3 - K_3 \hat{q}_2^2 + K_1 \hat{q}_1 (-1 + 2 \hat{q}_2^2) \hat{q}_3 \nonumber \\
&& + K_3 (-1 + 2 \hat{q}_2^2) \hat{q}_3^2) - \hat{q}_2 (14 K_1 K_3 \hat{q}_1 (1 - 2 \hat{q}_3^2) + 3 K_1^2 \hat{q}_3 (-19 + 28 \hat{q}_2^2 + 28 \hat{q}_3^2) + K_3^2 \hat{q}_3 (-43 + 70 \hat{q}_2^2 + 56 \hat{q}_3^2))) \nonumber \\
&& + 13 \ell^6 K^2 (-23 K_1^4 \hat{q}_2 \hat{q}_3 (-31 + 28 \hat{q}_2^2 + 28 \hat{q}_3^2) + 322 K_1^3 \hat{q}_1 (K_2 (-1 + 2 \hat{q}_2^2) \hat{q}_3 + K_3 \hat{q}_2 (-1 + 2 \hat{q}_3^2)) \nonumber \\
&& + 70 K_1 (K_2^2 + K_3^2) \hat{q}_1 (K_2 (-1 + 2 \hat{q}_2^2) \hat{q}_3 + K_3 \hat{q}_2 (-1 + 2 \hat{q}_3^2)) - 2 K_1^2 (K_3^2 \hat{q}_2 \hat{q}_3 (-230 + 287 \hat{q}_2^2 + 196 \hat{q}_3^2) \nonumber \\
&& + K_2^2 \hat{q}_2 \hat{q}_3 (-230 + 196 \hat{q}_2^2 + 287 \hat{q}_3^2) - 91 K_2 K_3 (1 - \hat{q}_2^2 + (-1 + 2 \hat{q}_2^2) \hat{q}_3^2)) - (K_2^2 + K_3^2) \nonumber \\
&& \times (K_2^2 \hat{q}_2 \hat{q}_3 (-251 + 252 \hat{q}_2^2 + 182 \hat{q}_3^2) + K_3^2 \hat{q}_2 \hat{q}_3 (-251 + 182 \hat{q}_2^2 + 252 \hat{q}_3^2) + 70 K_2 K_3 (1 - \hat{q}_2^2 + (-1 + 2 \hat{q}_2^2) \hat{q}_3^2))) \nonumber \\
&& + 7 \ell^4 (K_1^4 \hat{q}_2 \hat{q}_3 (2999 \hat{q}_1^2 - 25 (-2 + \hat{q}_2^2 + \hat{q}_3^2)) + 868 K_1^3 \hat{q}_1 (K_2 (-1 + 2 \hat{q}_2^2) \hat{q}_3 + K_3 \hat{q}_2 (-1 + 2 \hat{q}_3^2)) \nonumber \\
&& + 672 K_1 (K_2^2 + K_3^2) \hat{q}_1 (K_2 (-1 + 2 \hat{q}_2^2) \hat{q}_3 + K_3 \hat{q}_2 (-1 + 2 \hat{q}_3^2)) - (K_2^2 + K_3^2) (K_2^2 \hat{q}_2 \hat{q}_3 (-1481 + 1484 \hat{q}_2^2 \nonumber \\
&& + 1456 \hat{q}_3^2) + K_3^2 \hat{q}_2 \hat{q}_3 (-1481 + 1456 \hat{q}_2^2 + 1484 \hat{q}_3^2) + 28 K_2 K_3 (1 - \hat{q}_3^2 + \hat{q}_2^2 (-1 + 2 \hat{q}_3^2)))  \nonumber \\
&& + 2 K_1^2 (K_2^2 \hat{q}_2 \hat{q}_3 (2069 - 2058 \hat{q}_2^2 - 2142 \hat{q}_3^2) + K_3^2 \hat{q}_2 \hat{q}_3 (2069 - 2142 \hat{q}_2^2 - 2058 \hat{q}_3^2) \nonumber \\
&& + 84 K_2 K_3 (1 - \hat{q}_3^2 + \hat{q}_2^2 (-1 + 2 \hat{q}_3^2)))).
\end{eqnarray}
\end{widetext}
Note that $\mathcal{C}_{33}^{(n)}(\hat{\mathbf{q}},\mathbf{K})$ and $\mathcal{C}_{13}^{(n)}(\hat{\mathbf{q}},\mathbf{K})$ can be found from $\mathcal{C}_{22}^{(n)}(\hat{\mathbf{q}},\mathbf{K})$ and $\mathcal{C}_{12}^{(n)}(\hat{\mathbf{q}},\mathbf{K})$, respectively, by the transformations $K_2 \leftrightarrow K_3$ and $\hat{q}_2 \leftrightarrow \hat{q}_3$ (the source being along $1$), and that the matrix $W_{ij}(\mathbf{q},\mathbf{K})$ is symmetric. The behavior of $W_{ij}(\mathbf{q},\mathbf{K})$ in Eq.~(\ref{eq:two-point_field_correlation_complete_1}) at small $K$ is dominated by the first term of the sum, which is the only one that diverges.

\section{Diffusion of the energy density}\label{sec:A3}

In this Appendix, we show that the classical expression for the diffusion of the energy density is properly recovered using Eq.~(\ref{eq:fieldcorrelationbulk}). Starting from this equation and integrating over $\mathbf{q}$, we have
\begin{eqnarray}
\int \langle E_i && (\mathbf{q}+\frac{\mathbf{K}}{2}) E_j^\star(\mathbf{q}-\frac{\mathbf{K}}{2}) \rangle \frac{d\mathbf{q}}{(2\pi)^3} \nonumber \\
&& = \frac{1}{K^2 \ell^2} \int (\delta_{ij} - \hat{q}_i \hat{q}_j) \langle G(\mathbf{q}) \rangle \langle G^\star(\mathbf{q}) \rangle \frac{d\mathbf{q}}{(2\pi)^3}  \nonumber \\
&& = \frac{4 \pi}{K^2 \ell^2} \langle (\delta_{ij} - \hat{q}_i \hat{q}_j) \int \langle G(\mathbf{q}) \rangle \langle G^\star(\mathbf{q}) \rangle q^2 \frac{dq}{(2\pi)^3} \rangle_{\hat{\mathbf{q}}}   \nonumber \\
&& = \frac{1}{6 \pi K^2 \ell} \delta_{ij},
\end{eqnarray}
where we have used the results shown in Eq.~(\ref{eq:residue}) for the integral and $\langle \delta_{ij} - \hat{q}_i \hat{q}_j \rangle_{\hat{\mathbf{q}}} = \frac{2}{3}\delta_{ij}$. Transforming back the expression into real space, we reach
\begin{equation}
\langle E_i(\mathbf{R}) E_j^\star (\mathbf{R}) \rangle = \frac{1}{24 \pi^2 \ell R} \delta_{ij}=\frac{\langle |\mathbf{E}(\mathbf{R})|^2 \rangle}{3} \delta_{ij},
\end{equation}
where $\langle |\mathbf{E}(\mathbf{R})|^2 \rangle=\frac{1}{8 \pi^2 \ell R}$ is the averaged intensity throughout the medium. We finally recover the classical expression for the energy density in a disordered medium, that is the solution of the diffusion equation
\begin{equation}
U(\mathbf{R})=\frac{6 \pi}{c} \text{Tr}[\langle E_i(\mathbf{R}) E_j^\star (\mathbf{R}) \rangle]=\frac{1}{4 \pi \mathcal{D} R},
\end{equation}
where $\mathcal{D}=\frac{c \ell}{3}$ is the diffusion constant.

\section{Details of calculations for the two-point field correlation matrix}\label{sec:A4}

In this Appendix, we detail the inverse Fourier transform of the two-point field correlation matrix $W_{i,j}(\mathbf{q},\mathbf{K})$ in Eq.~(\ref{eq:fieldcorrelationbulk}). The inverse Fourier transform in $\mathbf{K}$ is straightforward and gives $\mathcal{F}^{-1}_\mathbf{K} \left( \frac{1}{K^2 \ell^2} \right)=\frac{1}{4 \pi \ell^2 R}$. For the inverse Fourier transform in $\mathbf{q}$, it is convenient to consider the two polarization-dependent terms, $\delta_{ij}$ and $-\hat{q}_i \hat{q}_j$, separately. The inverse Fourier transform can be simplified in spherical coordinates, with $\hat{q}=\cos(\theta) \hat{u}_1 + \sin(\theta) \cos(\phi) \hat{u}_2 + \sin(\theta) \sin(\phi) \hat{u}_3$, setting $\mathbf{X}=X \hat{u}_1$ without loss of generality, and considering that the averaged Green function does not depend on direction, $\langle G(\mathbf{q}) \rangle = \langle G(q) \rangle$. We find
\begin{eqnarray}\label{eq:integralfirstterm}
\int \delta_{ij} && \langle G(\mathbf{q}) \rangle \langle G^\star(\mathbf{q}) \rangle \exp[-i \mathbf{q} \cdot \mathbf{X}] \frac{d\mathbf{q}}{(2\pi)^3} \nonumber \\
&& = \frac{\delta_{ij}}{2\pi^2 X} \int_0^\infty q \sin(qX) \langle G(q) \rangle \langle G^\star (q) \rangle  dq,
\end{eqnarray}
for the first term and
\begin{eqnarray}\label{eq:integralsecondterm}
\int \hat{q}_i \hat{q}_j && \langle G(\mathbf{q}) \rangle \langle G^\star(\mathbf{q}) \rangle \exp[-i \mathbf{q} \cdot \mathbf{X}] \frac{d\mathbf{q}}{(2\pi)^3} \nonumber \\
&& = \frac{1}{2\pi^2 X^3} \int_0^\infty C_{ij} \langle G(q) \rangle \langle G^\star(q) \rangle dq,
\end{eqnarray}
for the second term, with
\begin{equation}\label{eq:Cij}
C_{ij}=
\begin{cases}
2X \cos(qX) + (q X^2 -\frac{2}{q}) \sin(qX) \quad  (i=j=1) \\
-X \cos(qX) + \frac{1}{q} \sin(qX) \quad  (i=j=2,3) \\
0 \quad  (i \neq j).
\end{cases}
\end{equation}

The correlation function of the different elements of $W_{ij}(\mathbf{R},\mathbf{X})$ can now be calculated analytically. Separating the different terms that are summed in these integrals, and defining $n_e=\sqrt{1+i/(k_0 \ell)}$, we get
%
\begin{eqnarray}\label{eq:integral1}
\frac{1}{2 \pi^2 X} && \int_0^\infty  q \sin(qX) \langle G(q) \rangle \langle G^\star(q) \rangle dq \nonumber \\
&& = \frac{-i}{4 \pi^2 X} \int_{-\infty}^\infty q \exp[iqX] \langle G(q) \rangle \langle G^\star(q) \rangle dq \nonumber \\
&& = \frac{-i \ell}{8 \pi k_0 X} \Big( \exp[ik_0 X n_e] - \exp[-ik_0 X n_e^\star] \Big), \nonumber \\
\end{eqnarray}
\begin{eqnarray}\label{eq:integral2}
\frac{1}{2 \pi^2 X^2} && \int_0^\infty  \cos (qX) \langle G(q) \rangle \langle G^\star(q) \rangle dq \nonumber \\
&& = \frac{1}{4 \pi^2 X^2} \int_{-\infty}^\infty \exp[iqX] \langle G(q) \rangle \langle G^\star(q) \rangle dq \nonumber \\
&& = \frac{\ell}{8 \pi k_0 X^2} \left[ \frac{\exp[-ik_0 X n_e^\star]}{k_0 n_e^\star} + \frac{\exp[ik_0 X n_e]}{k_0 n_e} \right] \nonumber \\
\end{eqnarray}
and
\begin{eqnarray}\label{eq:integral3}
\frac{1}{2 \pi^2 X^3} && \int_0^\infty  \frac{\sin(qX)}{q} \langle G(q) \rangle \langle G^\star(q) \rangle dq \nonumber \\
&& = \frac{-i}{4 \pi^2 X^3} \int_{-\infty}^\infty \frac{\exp[iqX]}{q} \langle G(q) \rangle \langle G^\star(q) \rangle dq \nonumber \\
&& = \frac{\ell^2}{8 \pi k_0^2 (1+ k_0 ^2 \ell^2) X^3} \nonumber \\
&& \times \Big[2 + i k_0 \ell n_e^2 \exp[-ik_0 X n_e^\star] \nonumber \\
&& - i k_0 \ell n_e^{\star2} \exp[ik_0 X n_e] \Big].
\end{eqnarray}
These integrals were solved by the residue theorem on the upper half-space and the latter exhibits a pole at $q=0$ that needs to be properly taken into account. Inserting Eqs.~(\ref{eq:integral1})-(\ref{eq:integral3}) into Eq.~(\ref{eq:Cij}) leads to the final expression for the two-point correlation function in Eq.~(\ref{eq:coherence}).

\bibliographystyle{apsrev}

\end{document}